\documentclass[showpacs,preprintnumbers]{revtex4}
\usepackage{graphicx}
\begin{document}
\title{On the chirality of quark modes}
\author{Alex C. Kalloniatis\footnote{akalloni@physics.adelaide.edu.au} }

\address{
Special Research Centre for the Subatomic Structure of Matter,
University of Adelaide,
South Australia 5005, Australia}

\author{Sergei N. Nedelko \footnote{nedelko@thsun1.jinr.ru}\footnote{A 
preliminary part of this work was done during a stay at the Institute of 
Theoretical Physics~III of
Erlangen-Nuremberg University, Erlangen, Germany.}}
\address{
Bogoliubov Laboratory of Theoretical Physics, JINR,
141980 Dubna, Russia}

\date{\today}
\preprint{ADP-02 80 T519}

\begin{abstract}
A model for the QCD vacuum based on a domainlike structured
background gluon field with definite duality attributed to the domains
has been shown elsewhere to give confinement of static
quarks, a reasonable value for the topological susceptibility and
indications that chiral symmetry is spontaneously broken. In this paper  
we study in detail the eigenvalue problem for the Dirac operator
in such a gluon mean field. A study of the local chirality parameter 
shows that the lowest nonzero eigenmodes
possess a definite mean chirality correlated with the duality of a 
given domain. 
A probability distribution of the local chirality qualitatively 
reproduces histograms seen in lattice simulations.
\end{abstract}
\pacs{12.38.Aw 12.38.Lg 14.70.Dj 14.65.Bt 11.15.Tk}

\maketitle
\section{Introduction}
In a previous paper  \cite{KaNe01} 
we formulated a model which characterises the QCD vacuum  
by a ``lumpy''  distribution of field strength and topological charge density.
For lack of a better name we shall refer to the model as ``the domain model''. 
The formulation is given concretely in terms of a partition function
which describes a statistical ensemble of domains, each of which is
characterised by a set of internal parameters associated with the 
mean background gluon field, and 
the internal dynamics are represented by fluctuation fields.
Correlation functions in this model can be calculated by 
taking the mean field into account 
explicitly and decomposing over the fluctuations.  
We briefly review the details and assumptions behind the model   
in the next section but state here unambiguously that the ``domains'' 
in question  are assumed to be purely quantum in nature. They   
are not semi-classical solutions of Yang-Mills theory
and are not argued to exist  
as topologically stable classical configurations,  
rather they seek to characterise the  average bulk properties
of the ensemble of fields that determine the gluonic vacuum.  
In particular, it is not assumed that the topological charge 
associated with a domain should be an integer.
The rationale of such an extremely simplified construction 
can be understood as an attempt to  implicitly
incorporate effects of the presence of 
singular pure gauge 
configurations in the QCD Euclidean functional integral into a
practical  calculational scheme with a mean field 
description of the QCD ground state.  
A self-consistent mean field approach requires
nonperturbative calculation of the free energy as a functional of the 
mean field whose minima should determine its form, but 
this is beyond the reach of analytical methods.  
Nevertheless there is an accumulation
of semi-qualitative arguments~\cite{KaNe01} in favour of the {\it ansatz} 
for the mean field we have chosen. 

In the gluonic sector the model depends
on two parameters, a mean field strength per domain $B$
and a mean size for domains $R$, which is sufficient for an adequate 
description of the pure glue characteristics of the QCD vacuum
-- the gluon condensate, topological susceptibility, and string tension.
The  Wilson loop in such a gluonic
background was found to exhibit an area law dependence for large
loops. Thus a confinement of static fundamental charges is captured
by the model; some dynamical gluon degrees of freedom turn out also
to be non-propagating.
The absolute value of underlying average topological charge per domain 
was determined to be approximately $q = 0.15$ and the density of 
domains to be as high as 42 fm$^{-4}$.
Although tentative signals of spontaneous chiral symmetry breaking 
were also obtained in \cite{KaNe01}, a more rigorous 
consideration of the fermionic spectrum and eigenmodes as well as
calculation of the 
quark determinant is required, which was missed in \cite{KaNe01}. 

In this paper we solve the eigenvalue problem for the
Dirac operator for the gluonic background and boundary conditions
adopted in the model and examine the chirality properties of the eigenmodes.
This is a necessary step for checking the status of chiral 
symmetry breaking in the domain model.
But in view of recent lattice results this problem is valuable also in its
own right.

There are strong hints in lattice Monte-Carlo simulations at 
intermediate-range structures in individual gluon configurations 
once fluctuations are filtered out by some means. 
For example, cooling or relaxation
algorithms are well established now \cite{cooling}, 
and can reveal instantonic like structures
after several sweeps of a given lattice configuration. However, as these
algorithms are designed precisely to locally minimise the action, it is
natural they should bring   objects with integer topological charge into
relief.  Alternately, and more relevant to the present work,
low-lying and zero-modes of the massless Dirac operator 
can be used as a probe of long-range gluonic structures
\cite{fermionmodes} although only recently did this become more reliable with
lattice fermions with good chirality properties. For example,   
exact index theorems are found to be satisfied on the lattice 
\cite{Zha02}
and the zero modes are seen to precisely correlate with instantonic structures
in the  raw lattice configuration, in the absence of cooling
\cite{EdH02}.
However, the exact zero modes of any finite volume simulation
cannot be those relevant to spontaneous
chiral symmetry breaking, rather the discrete spectrum of
low-lying non-zero modes should, in the infinite volume limit,
go over to a continuous band at zero, saturating the 
Banks-Casher relationship \cite{BaC80}. Such modes are sometimes
called ``pseudo zero modes''. 
Low-lying non-zero modes with strong signs of chirality
in regions of high action and topological charge densities would be a tool for
identification of the properties of  gluonic configurations relevant to chiral symmetry breaking.
Indeed, after an initial negative result \cite{Hor02}, recent results have
emerged showing precisely this: low-lying non-zero modes of the
overlap Dirac operator which seem to accrue to topological structures
and exhibit strong chirality, as measured by the
local parameter $X$ defined by
\begin{eqnarray}
\label{lch}
\tan \left( \frac{\pi}{4}(1 - X(x)) \right)
= {{|\psi_-(x)|} \over {|\psi_+(x)|} },
\end{eqnarray}
in regions where the probability density ${\psi}^{\dagger}(x) \psi(x)$
of these modes is maximal \cite{EdH02}.  
The verification of the instantonic nature of these objects
and their relevance to spontaneous chiral symmetry breaking in
the infinite volume is still being argued out in the literature
(see for example the recent large $N_c$ study of \cite{CTW02}).

An unbiased summary of the totality of available lattice results 
can be formulated as follows: that they support 
the importance of gluon configurations producing regions of
approximately ``locked'' chromo-electric/magnetic fields for 
chiral symmetry breaking
but do not yet confirm or rule out a specifically instantonic nature
for these configurations~\cite{Edw02}. 
A potential test which might clarify this would 
be a comparison of hadronic correlation functions between vacuum models
and lattice simulations. Such results are already available
for instanton-based models \cite{DeG01}.  A search for 
complementary scenarios for the vacuum consistent with lattice results  
and incorporating confinement (missed in the instanton models) 
is evidently timely. This paper is a step in that direction for
the domain-scenario.

The core of this paper is the Dirac eigenvalue/function problem 
for a spherical four-dimensional Euclidean
region of radius $R$ with bag-like boundary conditions on the fermions
and in the presence of a covariantly constant (anti-) self-dual gauge field 
\begin{eqnarray}
\label{evp-0}
&&\!\not\!D\psi(x)=\lambda \psi(x),
\nonumber\\
&&i\!\not\!\eta(x) e^{i\alpha\gamma_5}\psi(x)=\psi(x), \ x^2=R^2.
\end{eqnarray}
Here $\eta_\mu(x)=x_\mu/|x|$,  $D_\mu$ is 
the covariant derivative in 
the fundamental representation, 
\begin{eqnarray*}
&&D_\mu=\partial_\mu-i\hat B_\mu
=\partial_\mu + \frac{i}{2}\hat n B_{\mu\nu}x_\nu,
\\&&\hat n=t^an^a, \  \tilde B_{\mu\nu}=\pm  B_{\mu\nu},
\end{eqnarray*}
where the (anti-)self-dual tensor $B_{\mu\nu}$  is constant, and the 
Euclidean $\gamma-$matrices are in an anti-hermitean representation. 

The outcomes of this study are the peculiar chiral properties of 
eigenspinors $\psi(x)$:  there are no zero modes and
none of the modes is chiral but at the centre of a domain 
the local chirality parameter $X(x)$  is found to be
$$X(0)=\pm 1$$ 
for all modes with zero orbital momentum.
The sign of chirality and the duality of the tensor $B_{\mu\nu}$ are locked:
$(+1)\ -1$
for an (anti-)self-dual field.
Simultaneously the normal density for these modes is maximal at the centre.
At the boundary the local chirality $X$ is equal to zero for all modes. 
The chirality of the lowest mode is a 
monotonic function inside the region while
for the higher radial excitations the chirality alternates.  The detailed
form of $X(x)$ changes with the variation of an arbitrary
angle $\alpha$ in the boundary condition.  This angle is 
treated as a random variable. 
Calculating chiralities  averaged
over a small central region for the various lowest modes, and operating in the 
whole ensemble of domains,
we end up with a histogram  which represents the probability of 
finding a given smeared chirality among the set of lowest modes. The histogram 
qualitatively reproduces 
the lattice results for the chirality of low-lying Dirac modes
such as those of \cite{EdH02} and others.

After reviewing the
domain model in the next section,
we present details of the solution of the above-formulated problem
and then study the chirality properties of the
eigenmodes.
We conclude with a discussion and future prospects.
Technical details of calculations and conventions for this paper are 
relegated to the Appendices.

\section{Review of the model}

It has been suggested~\cite{Lenz} that the restrictive influence
of pure gauge singularities
(present in instanton, monopole and vortex configurations)
on surrounding quantum fluctuations
may be used for 
an approximate treatment of QCD dynamics. Due to the complex structure
of the manifold of gauge orbits in QCD, singular gauge fields
may be unavoidable in the course of the elimination of redundant variables.
Obstructions such as the
Gribov problem and condensation of monopoles are two examples of this
potentially more general statement.   
This has also long been advocated by van Baal \cite{Baal} in his
studies of the fundamental domain in small volume studies on the
torus and sphere. In particular, the proposal has been made
that ``domain formation'' at larger volumes can be driven essentially by 
the non-trivial topology of the gauge field manifold.
 Moreover, it is stressed in \cite{Baal} 
that the full set of singular fields, instantons, monopoles and 
vortices must play a role in this. 
One can also add to this  hierarchy domain wall singularities~
\cite{Ford98} 
which are not topologically stable on their own but can be part of a 
complicated object: a domain wall can start and end on a lower dimensional 
topologically nontrivial singularity of lower dimension, namely a vortex,
and in this sense should not be neglected also. 

An arbitrary gauge field configuration ${\cal A}$ containing 
a pure gauge singularity can be represented in the vicinity of the 
singularity as
\begin{eqnarray*}
{\cal A}_{\mu} = S_{\mu}+ Q_{\mu}
\end{eqnarray*}
with $S_{\mu}$ a pure gauge singular field. 
If we now substitute this into the Yang-Mills Lagrangian we will see that 
the requirement of finitness of the action density imposes specific conditions 
on the behaviour of $Q$ in the vicinity of the singularity in $S$. 
The model we consider focusses on domain wall singular hypersurfaces which 
are the most restrictive for $Q$;  an inclusion of lower dimensional singularities is a complicated task beyond the scope of the present work.
In the case of domain wall the constraining influence of $S$ on gluon fluctuations $Q$
and quark fields $\psi$ is expressed via the
boundary conditions 
\begin{eqnarray}
\left[ Q,S \right] =  0 \label{gluebc}, \\ 
\bar\psi(x)\!\not\!\eta(x)\psi(x)=0 
\label{genquarkbc} 
\end{eqnarray} 
for $x$ being on the singular hypersurface of the pure gauge field $S$.
These conditions ensure a non-vanishing weight for such fields in the 
functional integral.

Domain wall singular pure gauge configurations are topologically trivial.
This implies that the field $S$ can be characterised by a definite 
colour direction  $n^a$ and the matrix $n^at^a$  can always be tuned to 
belong to the Cartan subalgebra of $SU_{\rm c}(3)$. 
The off-diagonal (or, equivalently, orthogonal to $n^a$) 
components of the fluctuations 
$Q$ must then satisfy Dirichlet boundary conditions,
while those fluctuations longitudinal to $n^a$ are not restricted at the 
domain wall. A typical configuration of this type looks like a system  of  
domains which are coupled in a sense that fluctuations inside neighbouring  
domains interact with each other via exchange by the
gluon modes longitudinal to the colour direction of the domain boundaries.
It should be stressed that unavoidably there are obstructions of 
colour direction at the domain wall junctions where lower dimensional  
topologically nontrivial singularities are situated.

To be specific and to deal with an analytically tractable model  we 
introduce several drastic simplifications:
we disengage ourselves from the obstructions in the colour direction and
substitute the coupling between domains by the presence of a mean field.
Inside and on the boundary of the domain 
the field is taken to be covariantly constant  
(anti-)self-dual such that the  strength over the whole Euclidean space 
reads
\begin{equation}
 F^{a}_{\mu\nu}(x)
=\sum_{j=1}^N n^{(j)a}B^{(j)}_{\mu\nu}\theta(1-(x-z_j)^2/R^2), 
\ B^{(j)}_{\mu\nu}B^{(j)}_{\mu\rho}=B^2\delta_{\nu\rho}
\end{equation}
The individual colour and space orientations in each domain are random. 
In particular, effective action arguments were used in \cite{KaNe01}
to constrain the form of  $n^{(j)a}$ such that the matrix 
$\hat n^{(j)}=t^3\cos\xi_j+t^8\sin\xi_j$ 
with angles $\xi_j\in\{\frac{\pi}{6}(2k+1),\ k=0,\dots,5\}$ 
corresponding to the discrete Weyl subgroup.
Domains are taken to be hyperspherical with a mean radius $R$ and
centered at random points $z_j$.  
For a detailed motivation of these steps we refer the reader to \cite{KaNe01}.

In this way,  consideration is reduced to a model with 
essentially two free parameters: the mean field strength $B$ and the 
mean domain radius $R$. The partition function for this simplified system
can be written down as
\begin{eqnarray}
{\cal Z} & = & {\cal N}\lim_{V,N\to\infty}
\prod\limits_{i=1}^N
\int\limits_{\Sigma}d\sigma_i
\int_{{\cal F}_\psi^i}{\cal D}\psi^{(i)} {\cal D}\bar \psi^{(i)}
\int_{{\cal F}^i_Q} {\cal D}Q^i 
\delta[D(\breve{\cal B}^{(i)})Q^{(i)}]
\Delta_{\rm FP}[\breve{\cal B}^{(i)},Q^{(i)}]
e^{
- S_{V_i}^{\rm QCD}
\left[Q^{(i)}+{\cal B}^{(i)}
,\psi^{(i)},\bar\psi^{(i)}
\right]}
\label{partf}
\end{eqnarray}
where the functional spaces of integration
${\cal F}^i_Q$ 
and ${\cal F}^i_\psi$  are specified by the boundary conditions  
$(x-z_i)^2=R^2$
\begin{eqnarray}
\label{bcs}
&&\breve n_i Q^{(i)}(x)=0, 
\\
&&i\!\not\!\eta_i(x) e^{i\alpha_i\gamma_5}\psi^{(i)}(x)=\psi^{(i)}(x),
\label{quarkbc} \\
&&\bar \psi^{(i)} e^{i\alpha_i\gamma_5} i\!\not\!\eta_i(x)=-\bar\psi^{(i)}(x).
\label{adjquarkbc} 
\end{eqnarray}
Here $\breve n_i= n_i^a t^a$ with the colour 
generators $t^a$ in the adjoint representation. 
The conditions Eqs.(\ref{quarkbc}) and (\ref{adjquarkbc}) 
represent specific (though not unique) choices for the implementation
of Eq.(\ref{genquarkbc}) which manifest the explicit breaking of chiral
symmetry by the boundary condition, as occurs for example
in bag models for the nucleon. 
The thermodynamic limit assumes $V,N\to\infty$ but 
with the density $v^{-1}=N/V$ taken fixed and finite. The
partition function is formulated in a background field gauge
with respect to the domain mean field.
The measure of integration over parameters characterising domains is 
\begin{eqnarray}
\label{measure}
\int\limits_{\Sigma}d\sigma_i\dots & = & \frac{1}{48\pi^2}
\int_V\frac{d^4z_i}{V}
\sum_{\nu_i=-\infty}^\infty
\int_{(2\nu_i-1)\pi}^{(2\nu_i+1)\pi}d\alpha_i
\int\limits_0^{2\pi}d\varphi_i\int_0^\pi d\theta_i\sin\theta_i
\nonumber\\
&\times&\int_0^{2\pi} d\xi_i\sum\limits_{l=0,1,2}^{3,4,5}
\delta(\xi_i-\frac{(2l+1)\pi}{6})
\int_0^\pi d\omega_i\sum\limits_{k=0,1}\delta(\omega_i-\pi k)
\dots ,
\end{eqnarray}
where $(\theta_i,\varphi_i)$ are the spherical angles of the 
chromomagnetic field, $\omega_i$ is the angle between chromoelectric and 
chromomagnetic fields and $\xi_i$ is an angle parametrising the colour 
orientation. It should be noted that because of the axial anomaly and 
that nothing {\it a priori} constrains the topological charge per domain to be 
integral the fermion determinant is a single valued function of 
$\alpha_i$ only if an appropriate Riemman surface is constructed.
Here $\nu_i$  enumerates the Riemman sheets to be taken into account.

This partition function describes a statistical system 
of density $v^{-1}$ composed of extended
domain-like structures, each of which is
characterised by a set of internal parameters and
whose internal dynamics are represented by the fluctuation fields.
It respects all the symmetries of the QCD Lagrangian, since the statistical 
ensemble is invariant under space-time and colour gauge symmetries. 
For the same reason, if the quarks are massless then the
chiral invariance is respected.

Field eigenmodes satisfying the above boundary conditions
in the presence of an (anti-)self-dual gluon field and 
corresponding Green functions can be calculated explicitly.
For gluons, this was shown in \cite{KaNe01}. For quarks, this will
be shown in this paper.
On this basis one can compute any correlation function taking the mean field 
into account exactly and decomposing the integrand over fluctuations.
In particular, correlation functions of the mean field itself 
have a finite radius $R$, which is more or less obvious and 
is discussed in~\cite{KaNe01} in detail.

Within this framework the gluon condensate to lowest order in fluctuations
is immediately obtained in the form
\begin{equation}
g^2 \langle F^a_{\mu\nu}(x)F^a_{\mu\nu}(x)\rangle=4B^2,
\end{equation}
and the topological susceptibility reads
\begin{eqnarray*}
\chi = \int d^4x \langle Q(x) Q(0) \rangle 
= {{B^4 R^4} \over {128 \pi^2}}.
\end{eqnarray*}
Less trivial is the manifestation of an area law
for static quarks. Computation of the Wilson
loop for a circular contour  of
a large  radius $L\gg R$ gives a string tension $\sigma = B f(\pi B R^2)$
with the function 
\begin{eqnarray}
\label{stringsu3}
f(z)=\frac{2}{3 z}
\left(3-
\frac{\sqrt{3}}{2z}\int_0^{2z/\sqrt{3}}\frac{dx}{x}\sin x
- \frac{2\sqrt{3}}{z}\int_0^{z/\sqrt{3}}\frac{dx}{x}\sin x
\right)
\nonumber
\end{eqnarray}
Estimations of the values of these quantities are known from lattice calculation or phenomenological approaches and can be used to fit  $B$ and $R$.
As described in \cite{KaNe01} these parameters are fixed to
be
\begin{equation}
\sqrt{B} = 947 {\rm {MeV}}, R=(760 {\rm{MeV}})^{-1} = 0.26 {\rm {fm}}
\end{equation}
with the average absolute value of topological charge per domain
turning out to be $q\approx 0.15$ and the density of domains 
$v^{-1}=42{\rm fm}^{-4}$. The topological susceptibility then
turns out to be $\chi \approx (197 {\rm MeV})^4$, comparable to the 
Witten-Veneziano value \cite{largeNc}.

\section{Spectrum of the Dirac operator in a domain}

We have mentioned already that the boundary conditions
on fermions violate chiral symmetry explicitly which can only be
restored by a random assignment of values of $\alpha$
over the complete ensemble of domains in Euclidean space.

In this section we address the eigenvalue problem 
for the massless Dirac operator as it is stated in Eqs.~(\ref{evp-0}).
Here we give the scheme for solving the problem, with technical 
details given in the appendices. 
The Dirac matrices in Euclidean space are chosen to be anti-hermitean 
and taken in the chiral representation.

For boundary conditions on a hypersphere and covariantly constant
background field of definite duality
it is natural to use hyperspherical coordinates $(r,\Omega)$,
given in detail in Appendix A. In such coordinates,  
rather than work with the covariant derivative itself,    
it is more convenient to introduce the operator $\!\not\!\eta \!\not\!D$
which can be easily expanded into intrinsic and orbital angular momentum
generators. Any spinor can be represented in the form   
\begin{eqnarray}
\label{efr}
\psi=i\!\not\!\eta \chi + \varphi, \  \bar\psi=i \bar\chi \!\not\!\eta  + 
\bar\varphi,
\end{eqnarray}
where $\varphi$ and $\chi$ have  the same chirality. This is simply a decomposition into a sum of chiral components.
The eigenvalue equation Eq. (\ref{evp-0}) can be rewritten 
then identically as
\begin{eqnarray}
\label{main}
\chi  =  -\frac{1}{i\lambda} \!\not\!\eta \!\not\!D\!\varphi, \ 
\!\not\!D^2\varphi  =  \lambda^2 \varphi.
\end{eqnarray} 
In these terms the boundary conditions take the form
\begin{eqnarray}
\label{bc-main}
\chi=- e^{ \mp i\alpha}\varphi,    \  \bar\chi= \bar\varphi e^{ \mp i\alpha},
\ x^2=R^2,
\end{eqnarray}
where upper (lower) signs correspond to  $\varphi$ and $\chi$
with chirality $\mp1$. 

A solution of Eqs.~(\ref{main}) is achieved by separating 
the angular and radial coordinates. 
To do this one has to represent respectively ${\!\not\!D}^2$ 
and ${\!\not\!\eta} {\!\not\!D}$ in terms of momentum generators
and projectors onto the various spin and colour polarization subspaces.
In four-dimensional Euclidean space 
the angular momentum operators can be represented as
\begin{eqnarray*}
{\bf K}_{1,2} = \frac{1}{2} ({\bf L} \pm {\bf M})
\end{eqnarray*}
with ${\bf L}$ the usual three-dimensional angular momentum
operator and ${\bf M}$ the Euclidean version of the boost operator.
These correspond to the decomposition of the four dimensional
rotational group $SO(4)$ into a product of two $SO(3)$ groups. 
They lead to Casimirs and eigenvalues
\begin{eqnarray*}
{\bf K}_1^2 & = & {\bf K}_2^2 \rightarrow 
            \frac{k}{2}(\frac{k}{2} + 1), \  k=0,1,\dots,\infty \\
K^z_{1,2} & \rightarrow & m_{1,2}, \   m_{1,2}=-k/2, -k/2+1,\dots,k/2-1,k/2,
\end{eqnarray*}
and the correponding angular eigenfunctions $C_{k m_1 m_2}(\Omega)$, given 
explicitly in Appendix A, are
labelled by orbital momentum $k$ and two
azimuthal numbers $m_1$ and  $m_2$.
Eigenstates are also characterized by the colour-spin polarisation
related to the projectors
\begin{eqnarray}
O_{\pm} = N_+ \Sigma_{\pm} + N_- \Sigma_{\mp}
\end{eqnarray}
with 
\begin{eqnarray*}
N_{\pm} = \frac{1}{2}(1\pm \hat n/|\hat n|), \ 
\Sigma_\pm  = \frac{1}{2}(1\pm{ \bf\Sigma  B}/B)
\end{eqnarray*}
being respectively the separate projectors for colour and spin polarizations.
Below we denote the polarisation with respect to $O$ by $\kappa=\pm$.

 It is shown in Appendix B  that if 
the background field is (anti-)self-dual the boundary condition can only
be implemented if spinors ${\varphi}$ and $\chi$ are (right) left handed.
Also the presence of the homogeneous background field 
reduces the spherical symmetry of the problem down to an axial-symmetry.  
In the representation implemented here this manifests itself 
as a restriction on the values of one of the azimuthal quantum numbers, 
namely $m_2=\pm k/2$ for the self-dual case and $m_1=\pm k/2$ 
for the anti-self-dual one.  The sign in front of $k/2$ is correlated with 
the spin polarization of the state as seen in the explicit 
expressions for the eigenspinors below.

Thus for the self-dual case,
$\gamma_5 \varphi = - \varphi$, $\gamma_5 \chi= - \chi$
so that the eigenspinors in the self-dual field can be labelled 
as $\psi^{- \kappa}_{k m_1}(x)$, while in the anti-self-dual field they are
$\psi^{+ \kappa}_{k m_2}(x)$.
With details in Appendix B,
we simply write down here the result for the self-dual case:   
\begin{eqnarray}
\psi^{-\kappa}_{km_1}&=&i\!\not\!\eta\chi^{-\kappa}_{km_1}
+\varphi^{-\kappa}_{km_1},
\label{psikappa}\\
\chi^{-+}_{km_1}
&=&-(i\Lambda)^{-1}z^{(k+1)/2}e^{-z/2}
\left[
M\left(k+2-\Lambda^2,k+2,z\right)
\right.
\nonumber\\
&-&\left.
\frac{k+2-\Lambda^2}{k+2}M\left(k+3-\Lambda^2,k+3,z\right)
\right]
\left(
\begin{array}{c}
0 \\
0  \\
N_-  
 {\cal C}_{km_1\frac{k}{2}}(\Omega)\\
N_+
{\cal C}_{k m_1 -\frac{k}{2}}(\Omega) 
\end{array} \right)
\nonumber\\
\varphi^{-+}_{km_1}
&=&z^{k/2}e^{-z/2}
M\left(k+2-\Lambda^{2},k+2,z\right)
\left(
\begin{array}{c}
0 \\
0  \\
N_-
 {\cal C}_{km_1\frac{k}{2}}(\Omega)\\
N_+
{\cal C}_{km_1 -\frac{k}{2}}(\Omega) 
\end{array} \right),
\label{psi-+}\\
\chi^{--}_{km_1}
&=&  z^{(k+1)/2}e^{-z/2}
\frac{i\Lambda}{k+2}M\left(1-\Lambda^2,k+3,z\right)
\left(
\begin{array}{c}
0 \\
0  \\
N_+  
 {\cal C}_{km_1\frac{k}{2}}(\Omega) \\
N_-
{\cal C}_{k m_1 -\frac{k}{2}}(\Omega)
\end{array} \right)
\nonumber\\
\varphi^{--}_{km_1}
&=&z^{k/2}e^{-z/2}
M\left(-\Lambda^{2},k+2,z\right)
\left(
\begin{array}{c}
0 \\
0  \\
N_+
 {\cal C}_{km_1\frac{k}{2}}(\Omega) \\
N_-
{\cal C}_{km_1 -\frac{k}{2}}(\Omega) 
\end{array} \right),
\label{psi--}
\end{eqnarray}
where $M(a,b,x)$ is the confluent hypergeometric function and 
\begin{eqnarray*}
z=\hat Br^2/2, \  
\Lambda=\lambda/\sqrt{2\hat B}, \  \hat B=|\hat n|B.
\end{eqnarray*} 
The projectors $N_\pm$ act on the colour vectors which are
implicit in above equations.  The eigenfunctions $\psi^{+\kappa}_{km_2}$ 
for the anti-self-dual case are
obtained by the change $m_1\to m_2$
and the shift of nonzero elements of the angular part to 
the first two positions of the spinor. 

\begin{figure}
\includegraphics{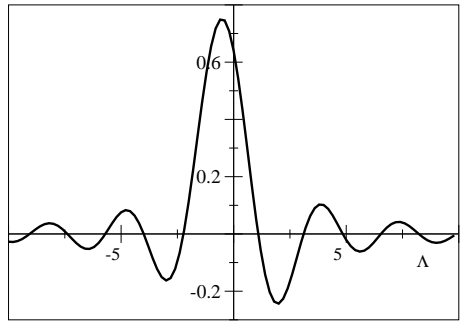}
\vspace*{14mm}
\caption{Graphical representation of the left hand side of
Eq.~(\ref{L--}) for k=0, 
a self-dual domain, and $\alpha=\pi/2$. Zeroes of the function are 
the radial eigenvalues $\Lambda^{--}_{0n}$.  }
\label{fig:Eqn}
\vspace*{6mm}
\end{figure}

The eigenvalues are determined  by the boundary condition
at $z=z_0 = \hat BR^2/2$ which for $\Lambda^{-+}_{k}$ takes the form
\begin{eqnarray}
\label{L-+}
e^{-i\alpha}M\left(k+2-\Lambda^2,k+2,z_0\right)
-\frac{\sqrt{z_0}}{i\Lambda}
\left[M\left(k+2-\Lambda^2,k+2,z_0\right)
-\frac{k+2-\Lambda^2}{k+2}M\left(k+3-\Lambda^2,k+3,z_0\right)
\right]=0,
\end{eqnarray}
and for $\Lambda^{--}_{k}$:
\begin{eqnarray}
\label{L--}
e^{-i\alpha}M\left(-\Lambda^2,k+2,z_0\right)+
\frac{i\Lambda \sqrt{z_0}}{k+2}
M\left(1-\Lambda^2,k+3,z_0\right)
=0.
\end{eqnarray}
The equations for the eigenvalues in an anti-self-dual domain
are the same as above but with $\alpha \rightarrow - \alpha$ as follows from 
Eqs.~(\ref{bc-main}).
The eigenvalues can be calculated numerically. They form a discrete set. 
Zero modes are absent, which is to be expected for these
types of boundary conditions~\cite{Wipf95}.
A graphical solution of Eq.~(\ref{L--}) at $\alpha=\pi/2$ 
is presented in Fig.\ref{fig:Eqn}  to illustrate the structure of the spectrum. 
In general, the eigenvalues are complex. 
The spectrum is real for  $\alpha=\pm\pi/2$ which is the only
value for which the boundary condition Eq.(\ref{bcs}) imposed on  
$\bar\psi^{\pm\kappa}_{km_{1,2}}$ is 
hermitean conjugated to the condition for $\psi^{\pm\kappa}_{km_{1,2}}$
and the general fermion field $\bar\psi$ can be decomposed in terms of 
the basis of conjugate  
eigenfunctions $\bar\psi^{\pm\kappa}_{km_{1,2}}$. For other values of 
$\alpha$ a biorthogonal basis should be introduced. In particular at $\alpha=0$
eigenvalues are complex and come in complex conjugated pairs.
The partition function is nevertheless real since if
$\lambda_{\rm sd}(\alpha)$ is an eigenvalue for the self-dual case 
then for the anti-self-dual domain there is 
an eigenvalue $\lambda_{\rm asd}(\alpha)$ such that
 \begin{eqnarray}
\label{sd-asd}
 \lambda_{\rm asd}(\alpha)=-\lambda^*_{\rm sd}(\alpha).
 \end{eqnarray}
We stress that the definition of $\lambda$ here does not include an 
imaginary unity in front of Dirac operator.

As seen from Fig.\ref{fig:Eqn}, in contradistinction to the eigenvalue problem 
in infinite volume on the space of square integrable functions the spectrum 
is not symmetric under reflections $\lambda\to-\lambda$.
This comes from the fact that $\gamma_5$ does not commute with the 
boundary condition so that $\gamma_5 \psi$ is not an eigenfunction if
$\psi$ is an eigenfunction. An  assymmetry of the spectrum is typical for the
Dirac operator in odd dimensional spaces (see \cite{Des98} and references
therein) and has important consequences there for the effective action. 
In our case
the unusual boundary conditions are responsible for the assymmetry in
 four dimensional Euclidean space~\cite{Wipf95}.

The most interesting 
feature of the fermionic eigenmodes becomes manifest if one 
considers the local chirality $X(x)$ of the lowest eigenmodes
as defined by Eq.~(\ref{lch}).

\begin{figure}[htb]
\vspace{17mm}
\includegraphics{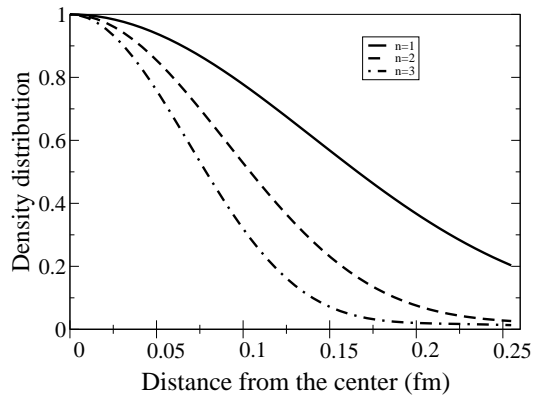}
\caption{Normal density distribution for modes with $k=0$ and $n=1,2,3$.}
\label{fig:k0n123}
\end{figure}

\section{Chirality of low lying modes}

\begin{figure}
\includegraphics{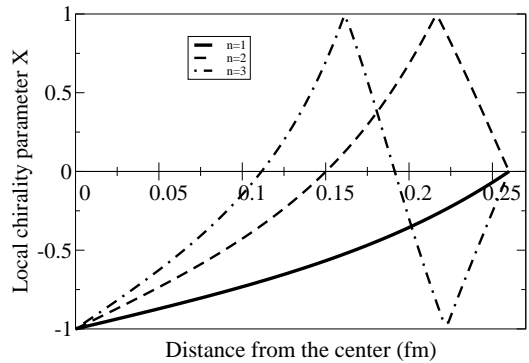}
\caption{Chirality parameter for the three lowest radial modes
$\psi_{00}^{--}$, self-dual domain, $\alpha=\pi/2$. }
\label{fig:X}
\vspace*{6mm}
\end{figure}

\begin{figure}[htb]
\vspace{17mm}
\includegraphics{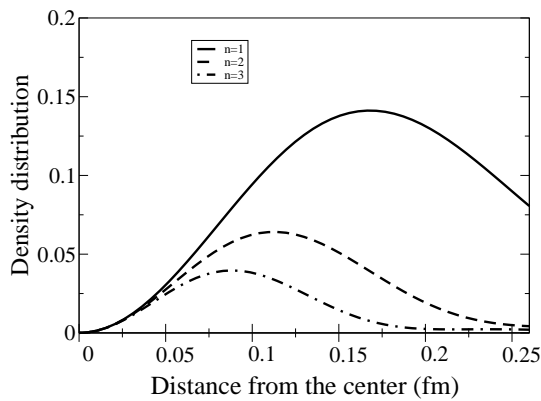}
\caption{Plot of the radial dependence of the 
normal density distribution for the modes with 
$k=1$ and $n=1,3, 5$.}
\label{fig:k1n135}
\end{figure}

It is obvious that none of the solutions are eigenstates of $\gamma_5$.  
However at the domain center $x^2=0$  (or $(x-z_j)^2=0$ in general) 
all the purely radial modes  with $k=0$ have a maximum in the 
probability density, they are chiral
and the sign of their chirality is determined 
by the duality of the mean field in a domain which is illustrated in
 Figs.{\ref{fig:k0n123}} and {\ref{fig:X}}.  
The probability density naturally vanishes at the domain center 
for the modes with $k>0$ as is seen in  Fig.{\ref{fig:k1n135}}.
To demonstrate this analytically let us turn to
the local chirality parameter given in the introduction which we rewrite here
in more detailed form
\begin{eqnarray*}
\tan \left( \frac{\pi}{4}(1 - X(x)) \right)
= \sqrt{{\psi^\dagger(x)(1-\gamma_5)\psi(x)} 
\over {\psi^\dagger(x)(1+\gamma_5)\psi(x)} }, \  -1\le X(x)\le 1,
\end{eqnarray*}
which takes the extremal values $X=\pm 1$  at positions
$x$ where  $\psi(x)$ is purely right(left) handed. 
Because $\varphi$ and $\chi$ have the same chirality, 
the representation Eq.(\ref{psikappa}) immediately gives
for the self-dual domain 
\begin{eqnarray*}
\tan \left( \frac{\pi}{4}(1 - X_{\rm sd}^{-\kappa}(x)) \right)
= {{|\varphi_{00}^{-\kappa}(x)|} 
\over {|\chi_{00}^{-\kappa}(x)|}  }, 
\end{eqnarray*}
while for the anti-self-dual case the local chirality reads 
\begin{eqnarray*}
\tan \left( \frac{\pi}{4}(1 - X_{\rm asd}^{+\kappa}(x)) \right)
= {{|\chi_{00}^{+\kappa}(x)|} 
\over {|\varphi_{00}^{+\kappa}(x)|}  }.
\end{eqnarray*}
Moreover due to the relation Eq.(\ref{sd-asd}) 
\begin{eqnarray*}
|\varphi_{00}^{+\kappa}(x)|=|\varphi_{00}^{-\kappa}(x)|, \
|\chi_{00}^{+\kappa}(x)|=|\chi_{00}^{-\kappa}(x)|.
\end{eqnarray*}
Representations Eqs.~(\ref{psi-+}) and (\ref{psi--}) show that
\begin{eqnarray*}
0<\lim_{x^2\to0}|\varphi_{00}^{+\kappa}(x)|<\infty, \
\lim_{x^2\to0}|\chi_{00}^{+\kappa}(x)|=0,
\end{eqnarray*}
which finally results in 
\begin{eqnarray*}
X_{\rm sd}^{+\kappa}(0)=-1,
\ 
X_{\rm asd}^{+\kappa}(0)=1.
\end{eqnarray*}
The local chirality parameter $X$
as a function of distance from the domain centre
for the lowest few modes is plotted in Fig.\ref{fig:X}.
There is a peak in $X$ at the domain centre. 
Away from the centre  $X$  decreases due to a competition of left and right
components of the eigenmodes  as the $\chi$ component
becomes non-vanishing.
As is seen from Fig.\ref{fig:X} the chirality of the lowest mode ($n=1$)
monotonically decreases with distance from the centre.
The chirality parameter for the excited modes alternates  
between extremal values,
the number of alternations is correlated with the radial number $n$, and the 
half-width decreases with growing $n$. The chirality parameter $X$
is zero at the boundary for all  modes. 
Qualitatively this picture does not depend on the angle $\alpha$.
The ``width'' of the peaks at half-maximum for the lowest ($n=0$) radial 
modes varies for different values $\alpha$ and is of the order
of $0.12-0.14 \ {\rm{fm}}$ if the values of $B$ and $R$ are fixed
from the gluon condensate and the string tension, 
consistent with the lattice observations of
\cite{Hor2}.

\begin{figure}[htb]
\vspace{17mm}
\includegraphics{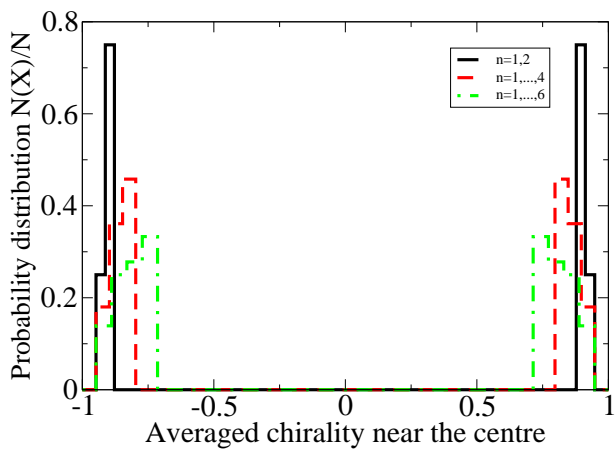}
\caption{Histogram of chirality parameter $\bar X$
averaged over the central region with radius $0.025$fm.
Plots given in solid, dashed and dot-dashed lines incorporate all modes with
$n\le2$, $n\le4$ and  $n\le6$ respectively.}
\label{fig:X-hist}
\end{figure}

We can now study the chirality characteristics of
the ensemble of fermion fields entering the partition
function Eq.(\ref{partf}) with all values of $\alpha$
treated with equal probability  
consistent with an explicit chiral symmetry. On the lattice
\cite{Hor02,EdH02} peaks in $X$ or $\psi^{\dagger} \psi$ 
would only be localiseable
within a size corresponding to the lattice spacing. To take this into account,
 we average $X(x)$ over a small neighbourhood of the domain centre.
Thus we compute the probability to find a given value
of $\bar X$, the smeared $X$, among the chiralities for the lowest modes.
The result given in Fig.\ref{fig:X-hist}
was obtained for three sets of modes: with $n\le 2,k=0$ (solid line),  
$n\le 4,k=0$ (dashed line) and  $n\le 6,k=0$ (dot-dashed line), and all possible
values of $\alpha$ and spin-colour polarizations. 
The solid line, formed from the lowest modes,
evidently indicates two narrow peaks with 
$\bar X\approx \pm 0.87$. This double peaking is not unexpected in view of the 
above discussed chirality properties.  
Including higher modes broadens the peaks and shifts their maxima.
This feature as well as the above mentioned values for the half-width and 
the density of domains is 
in qualitative and quantitative agreement with 
recent lattice results \cite{EdH02,Hor2}.   
It should be stressed that orbital excitations ($k > 0$) are not included in 
the histograms.
because the probability density for orbital modes vanishes at the centre.  
However there are maxima in the probability density  for these modes 
in peripheral regions of the domain.
The  local chirality $X$ is  significantly smaller in peak value
than those for the radial modes at the centre. 
Inclusion of orbital modes will broaden the peaks more 
and build up the central plateau.

\section{Discussion and conclusions}
The statement that signals for
spontaneous chiral symmetry breaking should be identifiable
in the specific chirality
properties of fermionic eigenmodes for some ``dominant''
gluonic background field is generally accepted.
Such signals are now being seen on the lattice, but nevertheless
there are not so many analytically 
explicit examples of this relationship available.
Instanton motivated models are certainly the most advanced example of 
this kind. But one aspect of instanton models lies in the 
`dual' use to which the exact fermion zero modes of the 
classical instanton background are subject:  
these modes have definite chirality
and do not contribute in the infinite volume limit, but are
used to model pseudo zero modes which should 
characterise the non-semi-classical system of strongly overlapping 
instantons and anti-instantons.

We have studied the spectrum of quark modes in a domainlike
structured gluon background field. 
Such a background is argued to 
characterise the bulk average properties of the vacuum in the presence
of strong intermediate range fluctuations and is not the result of a
semi-classical approximation. 
The spectrum exhibits definite chirality
properties. In particular, there are no zero modes because
of the conditions which fermion fields must satisfy on the boundaries of 
domains. Nonetheless
at the centre of domains all radial modes are purely chiral
and the sign of their chirality depends on whether the underlying
gluon field is self-dual or anti-self-dual. Moreover, the
sign of chirality at the centre persists over 
the whole domain for the lowest modes. 
Studying the local chirality parameter $X$ in a
chirally symmetric ensemble of domains  we obtain qualitatively similar
results to those seen in lattice calculations. 
We stress that this comparison with lattice results takes place
at the level of an ensemble of configurations not on a configuration
by configuration basis.

Insofar as these lattice results for chirality are argued 
as supporting the evidence for spontaneous chiral symmetry breaking,
the same can be said of the domain model. We note the absence of any 
explicit zero modes in achieving this. Namely, the
range of configurations needed to produce the types of effects
seen on the lattice are not restricted to instanton-like fields.
It suffices that a given gluon background admit strongly chiral low-lying  
non-zero modes. In this respect, the more significant 
property of the gluon background is the ``locking'' of
chromoelectric and chromomagnetic fields into self-dual or
anti-self-dual fields in relatively large but finite regions of space 
restricted by the 
hypersurfaces on which a pure gauge singularities are assumed to be situated.
It should be stressed that in the thermodynamic limit the number of domains 
is growing but their sizes stay fixed around some finite mean value.

The solutions obtained in this paper provide a basis for computation of
chiral condensate $\langle \bar{\psi} \psi \rangle$, in particular
in the presence of an explicit CP violating $\theta$ term.
In view of the 
 assymmetry of the spectrum, the specific chiral
properties of the eigenmodes and 
noninteger mean topological charge associated with a domain
are expected to simultaneously enhance 
the role of the axial anomaly in the 
spontaneous breakdown of chiral flavour symmetry and factor in the solution
of the $U_{\rm A}(1)$ problem
\cite{Des98,Cre78,Dash71}.

\section*{Acknowledgements}
SNN would like to acknowledge many fruitful discussions with Frieder Lenz
and Jan Pawlowski, and to thank the members of the Institute of 
Theoretical Physics III of the University of Erlangen-Nuremberg
for their kind hospitality.
ACK thanks Martin Oettel, Andreas Schreiber and Max Lohe
for numerous helpful discussions.
This work was partially supported by the grant of RFBR No.~01-02-17200
and the Australian Research Council.

\appendix 
  
\section{Conventions}

We use a chiral representation for the antihermitian Dirac matrices in
four Euclidean space,
\begin{eqnarray*}
&&\{\gamma_\mu,\gamma_\nu\}=-2\delta_{\mu\nu},  \  \gamma_\mu^+=-\gamma_\mu,
\\
&&\gamma_i =  \left( \begin{array}{cc}
       0 & \sigma_i \\
      -\sigma_i & 0 \end{array} \right), \ 
\gamma_4  =  i\left( \begin{array}{cc}
       0 & \bf{1} \\
      \bf{1} & 0  \end{array} \right),
\\
&&\gamma_5  =  \gamma_1\gamma_2\gamma_3\gamma_4 
         =  {\rm diag}(1,1,-1,-1).
\end{eqnarray*}
The background field is specified as 
\begin{eqnarray*}
&&\hat B_\mu(x)=-\frac{1}{2}\hat n B_{\mu\nu}x_\nu
\\
&&B_{ij}=\varepsilon_{ijk}B_{k}, \ B_{i}=\frac{1}{2}\varepsilon_{ijk}B_{jk},
\
 E_i=B_{i4}=\pm B_i,
\\
&&B_{\mu\nu}B_{\mu\rho}=\delta_{\nu\rho}B^2, \
\tilde B_{\mu\nu}=
\frac{1}{2}\varepsilon_{\mu\nu\alpha\beta}B_{\alpha\beta}=\pm B_{\mu\nu},
\end{eqnarray*}
In addition the following conventions and relations have been used
\begin{eqnarray*}
&&\sigma_{\mu\nu}=\frac{1}{2i}[\gamma_\mu,\gamma_\nu], 
\  \Sigma_i =\frac{1}{2}\varepsilon_{ijk}\sigma_{jk},
\\
&&\sigma_{ij}B_{ij}=2\Sigma_iB_i, \
\Sigma_\pm=\frac{1}{2}\left(1\pm\frac{\Sigma_iB_i}{B}\right)
\\
&&\sigma_{i4} =-\frac{1}{2}\gamma_5\varepsilon_{i4\mu\nu}\sigma_{\mu\nu}
=-\frac{1}{2}\gamma_5\varepsilon_{ijk}\sigma_{jk}=-\gamma_5\Sigma_i,
\\
&&\gamma_5\sigma_{\alpha\beta} = -\tilde\sigma_{\alpha\beta},
\
{\rm Tr}\gamma_\mu\gamma_\nu\gamma_\alpha\gamma_\beta\gamma_5=
4\varepsilon_{\mu\nu\alpha\beta},
\end{eqnarray*}
and in particular
\begin{eqnarray*}
\sigma_{\alpha\beta}B_{\alpha\beta}
&=&\sigma_{ij}B_{ij}+2\sigma_{i4}B_{i4}
=
2\Sigma_iB_i \mp 2B_i\gamma_5\Sigma_i
\\
&=&4P_\mp\Sigma_iB_i=
4BP_\mp(\Sigma_+ - \Sigma_-).
\end{eqnarray*}
We use the following hyperspherical coordinate system in $R^4$ 
\begin{eqnarray}
&&x_1 =  r \sin \eta \cos \phi, \ 
x_2  =  r \sin \eta \sin \phi, \nonumber \\
&& x_3  =  r \cos \eta \cos \chi, \ 
x_4 =  r \cos \eta \sin \chi ,
\label{sphcoords}
\end{eqnarray}
and define the angular momentum operators as follows
\begin{eqnarray}
L_i & = & -i \epsilon_{ijk} x_j \partial_k,
\nonumber \\
M_i & = & -i (x_4 \partial_i - x_i \partial_4),
\nonumber
\end{eqnarray}
respectively for spatial rotations and Euclidean ``boosts''.
As mentioned, it is more convenient to work in the basis
$$
{\bf K}_{1,2}  =  ({\bf L} \pm {\bf M})/2
$$
which generates the following Lie algebra 
\begin{eqnarray*}
\left[ K_1^i, K_1^j \right]  =  i \epsilon^{ijk} K_1^k , \
\left[ K_2^i, K_2^j \right]  =  i \epsilon^{ijk} K_2^k , \
\left[ K_1^i, K_2^j \right]  =  0.
\end{eqnarray*}
Thus the ladder operators 
\begin{eqnarray*}
K_{1,2}^{\pm}=(K_{1,2}^1\pm i K_{1,2}^2),
\end{eqnarray*}
satisfy the algebra
\begin{eqnarray*}
\left[ K_{1,2}^3, K_{1,2}^{\pm} \right] = \pm K_{1,2}^{\pm}
\end{eqnarray*}
and correspond to raising and lowering operators of $m_1, m_2$.

The angular eigenfunctions corresponding to the ${\bf K}_{1,2}$
generators are 
\begin{eqnarray*}
C_{k m_1 m_2}(\eta,\phi,\chi) & = &
(-1)^{|m_1 + m_2|} (2\pi)^{-1} \Theta_k^{m_1-m_2,m_1+m_2}(\eta),
\\
& \times& \exp i[(m_1-m_2)\chi + (m_1 +m_2)\phi] \nonumber \\
\Theta_k^{k-r-s,s-r}(\eta) & = & \sqrt{2(k+1)(k-r)!(k-s)! r! s!}
\\
& \times& \sum_{n=0}^r
{{ (-1)^{r-n} \cos^{k-r-s+2n} \eta \sin^{r+s-2n}\eta }
\over
{(k-r-s+n)! n! (r-n)!(s-n)!} },
\\
&&s=(k+m_2)/2, \ r=(k-m_1)/2,
\end{eqnarray*}
where $k, m_1, m_2$ are respectively the orbital angular
momentum and the two azimuthal quantum numbers, relevant for
an $O(4)=O(3)\times O(3)$ symmetry.
They take the following values
\begin{eqnarray*}
k =  0,1,2, \dots, \
m_1,m_2  =  -\frac{k}{2} , \dots, \frac{k}{2}.
\end{eqnarray*}

\section{Dirac eigenvalue problem in a domain.}

Here we give further details of the solution of
Eq.(\ref{evp-0}). 
Using the notation given in the main body, we can decompose
the field $\varphi$ over a  set of chiral and colour-spin projectors
$$\varphi=P_\pm\Phi_{0}+P_\mp O_+\Phi_{+1}+ P_\mp O_-\Phi_{-1},$$
where 
(lower)upper signs correspond to the (anti-)self-dual field 
background field, and 
fields $\Phi_{\zeta}$ must  satisfy the second order equation
\begin{eqnarray}
\label{soeq}
(-D^2+2\zeta\hat B -\lambda^2)P_\mp O_\zeta\Phi_\zeta=0.
\end{eqnarray}
We remind that implicitly $\Phi_\zeta^\alpha$ is the color vector  in the 
fundamental representation. 

If we were solving the problem 
for square integrable eigenfunctions in infinite volume
then all three components $\Phi_\zeta$
would enter the final set of eigenfunctions, moreover  as
is seen from  Eq.(\ref{soeq}), the  equation for the component $\Phi_{-1}$ 
would produce zero modes with chirality $\mp 1$. The spectrum in this case
would be discrete and all nonzero eigenvalues of Dirac operator 
come in pairs: if $\psi$ is an eigen function
with eigenvalue $\lambda$ then $\gamma_5\psi$ is an eigenfunction with 
eigen value $-\lambda$ -- pretty standard state of affairs.

However the bag-like boundary conditions Eq.(\ref{quarkbc})
we must satisfy in the present 
case change the structure of eigenfunctions and eigenvalues drastically. 
First of all, because of the identities
\begin{eqnarray*}
&&\gamma_\mu B_{\mu\rho}x_\rho P_\mp\Sigma_+=iB\!\not\!x P_\mp\Sigma_+,
\\
&&\gamma_\mu B_{\mu\rho}x_\rho P_\mp\Sigma_-=-iB\!\not\!x P_\mp\Sigma_-.
\end{eqnarray*}
and
\begin{eqnarray*}
&&\gamma_\mu B_{\mu\rho}x_\rho  P_{\pm}\Sigma_+
=\left\{\frac{2i}{x^2}B[ x_i^2-(B_ix_i)^2/B^2]-iB\right\}
\!\not\!x P_{\pm}\Sigma_+
\\
&&\gamma_\mu B_{\mu\rho}x_\rho  P_{\pm}\Sigma_-
=-\left\{\frac{2i}{x^2}B[ x_i^2-(B_ix_i)^2/B^2]-iB\right\}
\!\not\!x P_{\pm}\Sigma_-,
\end{eqnarray*}
which can be straightforwardly derived
by expanding both sides over a complete set of Dirac matrices,
the  bag-like boundary condition can be satisfied only for the trivial
solution  
$\Phi_0(x)\equiv0$. 
The significance of this observation
is that for (anti-)self-dual domains the boundary condition
can only be implemented on eigenspinors $\psi=i\!\not\!\eta \chi + \phi$ 
for (positive) negative chirality $\varphi$ and $\chi$. 
The function $\psi$ in turn is not an eigenspinor of $\gamma_5$,
which is natural because
the boundary condition violates chiral symmetry.  
Furthermore, zero modes are removed from the spectrum
because they must be chiral  but this is forbidden by boundary conditions.
And, finally, if $\psi$ is an eigenfunction
with eigenvalue $\lambda$ then $\gamma_5\psi$ is not an eigenfunction any more,
and there is no eigenvalue $-\lambda$ in the spectrum.

In order to find equations for components  $\Phi^\alpha_\zeta$ of the 
corresponding spinors we use that
\begin{eqnarray}
&&\varphi_{-\pm}=
P_-O_{\pm}\Phi_{\pm1} = \left(0,0,N_\mp\Phi^3_{\pm1},N_{\pm}\Phi^4_{\pm1}\right)^{\rm T},
\nonumber\\
&&\varphi_{+\pm}=
P_+O_{\pm}\Phi_{\pm1} = \left(N_\mp\Phi^1_{\pm1},N_{\pm}\Phi^2_{\pm1},0,0\right)^{\rm T}.
\end{eqnarray}
In hyperspherical coordinates Eqs.(\ref{sphcoords})
the equations for the spinor components read
(here and below we write down equations for the self-dual case only)
\begin{eqnarray}
&&\left\{
-\left[
\frac{1}{r^3}\partial_rr^3\partial_r - \frac{4}{r^2}{\bf K}_1^2
+2\frac{\hat n}{|\hat n|}\hat BK_{2z}-\frac{1}{4}\hat B^2r^2
\right]+2\hat B -\lambda^2
\right\}N_-\Phi_{+1}^3=0,
\nonumber\\
&&\left\{
-\left[
\frac{1}{r^3}\partial_rr^3\partial_r - {\frac{4}{r^2}}{\bf K}_1^2
+2\frac{\hat n}{|\hat n|}\hat BK_{2z}-\frac{1}{4}\hat B^2r^2
\right]+2\hat B -\lambda^2
\right\}N_+\Phi_{+1}^4=0,
\nonumber\\
&&\left\{
-\left[
\frac{1}{r^3}\partial_rr^3\partial_r - \frac{4}{r^2}{\bf K}_1^2
+2\frac{\hat n}{|\hat n|}\hat BK_{2z}-\frac{1}{4}\hat B^2r^2
\right]-2\hat B -\lambda^2
\right\}N_+\Phi_{-1}^3=0,
\nonumber\\
&&\left\{
-\left[
\frac{1}{r^3}\partial_rr^3\partial_r - \frac{4}{r^2}{\bf K}_1^2
+2\frac{\hat n}{|\hat n|}\hat BK_{2z}-\frac{1}{4}\hat B^2r^2
\right]-2\hat B -\lambda^2
\right\}N_-\Phi_{-1}^4=0,
\end{eqnarray}
The anti-self-dual case is reconstructed by the change $K_{2z}\to K_{1z}$
and $\Phi_{\zeta}^3\to \Phi_{\zeta}^1$, $\Phi_{\zeta}^4\to \Phi_{\zeta}^2$.
In \cite{KaNe01} we derived the general solution 
for equations of this type.  
The requirement of regularity at the origin then gives
\begin{eqnarray}
\Phi_{+1}^{3,km_1m_2}&=&N_-z^{k/2}e^{-z/2}
M\left(\frac{k}{2}+m_{2}-\Lambda^2+2,k+2,z\right)
 {\cal C}_{km_1m_2}(\varphi,\chi,\eta),
\nonumber\\
\Phi_{+1}^{4,km_1m_2}&=&N_+
z^{k/2}e^{-z/2}
M\left(\frac{k}{2}-m_{2}-\Lambda^2+2,k+2,z\right)
{\cal C}_{km_1m_2}(\varphi,\chi,\eta),
\nonumber\\
\Phi_{-1}^{3,km_1m_2}&=&N_+
z^{k/2}e^{-z/2}
M\left(\frac{k}{2}-m_{2}-\Lambda^2,k+2,z\right)
{\cal C}_{km_1m_2}(\varphi,\chi,\eta),
\nonumber\\
\Phi_{-1}^{4,km_1m_2}&=&N_-
z^{k/2}e^{-z/2}
M\left(\frac{k}{2}+m_{2}-\Lambda^2,k+2,z\right)
{\cal C}_{km_1m_2}(\varphi,\chi,\eta),
\nonumber
\end{eqnarray}
where $\Lambda=\lambda/\sqrt{2\hat B}$, $z=\hat Br^2/2$.
Thus the two independent mutually orthogonal solutions are
\begin{eqnarray}
\label{b4}
\varphi^{-+}&=&
\left( 
\begin{array}{c}
0 \\
0  \\
N_-
z^{k/2}e^{-z/2}
M\left(\frac{k}{2}+m_{2}-\Lambda^2+2,k+2,z\right)
 {\cal C}_{km_1m_2}\\
N_+
z^{k'/2}e^{-z/2}
M\left(\frac{k'}{2}-m'_{2}-\Lambda^{\prime 2}+2,k'+2,z\right)
{\cal C}_{k'm'_1m'_2}
\end{array} \right),
\\
\label{b5}
\varphi^{--}&=&
\left( 
\begin{array}{c}
0 \\
0  \\
N_+
z^{k/2}e^{-z/2}
M\left(\frac{k}{2}-m_{2}-\Lambda^2,k+2,z\right)
{\cal C}_{km_1m_2}\\
N_-
z^{k'/2}e^{-z/2}
M\left(\frac{k'}{2}+m'_{2}-\Lambda^{\prime 2},k'+2,z\right)
{\cal C}_{k'm'_1m'_2}
\end{array} \right),
\end{eqnarray}
where ``prime'' indicates that angular quantum numbers and eigenvalues 
in the third line need not
coincide with those in the fourth in order that these spinors be
eigenmodes of Eq.(\ref{main}).

To obtain an explicit representation for $\chi$ we use the identity:
\begin{eqnarray}
-\!\not\!\eta(x) \!\not\!D
=\partial_r
+2R^{-1}({\bf\Sigma}\cdot{\bf K}_1 P_+ +{\bf\Sigma}\cdot{\bf K}_2P_-)
-i\!\not\!\eta\frac{\hat n}{2}\gamma_\mu B_{\mu\nu}x_\nu,
\end{eqnarray}
where the action of the last term on $P_{\mp} O_{\zeta} \Phi^{\zeta}$ can
be determined via the identity
\begin{eqnarray} 
\frac{\hat n}{2}\gamma_\mu B_{\mu\nu}x_\nu P_\mp O_\zeta=
i\zeta\!\not\!\eta \frac{\hat BR}{2}P_\mp O_\zeta,
\nonumber
\end{eqnarray}
and the action of the ${\bf \Sigma} \cdot {\bf K}$ terms via  
\begin{eqnarray}
{\bf\Sigma}\cdot{\bf K}_{1,2}\sum_\zeta O_\zeta\Phi_\zeta & = & 
\left(
\Sigma_3 K^z_{1,2} + \Sigma^{(+)} K^-_{1,2} +\Sigma^{(-)} K^+_{1,2})
\right)\sum_\zeta O_\zeta\Phi_\zeta
\nonumber\\
   {} &=& 
\frac{\hat n}{|\hat n|}K^z_{1,2}\left(O_+\Phi_{+1}  -  O_-\Phi_{-1}\right)
\nonumber\\
&&+N_+\Sigma^{(+)}K^-_{1,2}\Phi_{+1}     + N_-\Sigma^{(+)}K^-_{1,2}\Phi_{-1} 
+ N_-\Sigma^{(-)}K^+_{1,2}\Phi_{+1} + N_+\Sigma^{(-)}K^+_{1,2}\Phi_{-1}.
\end{eqnarray}
As well as the ladder operators $K^{\pm}$ we also have analogous
operators for the spin, 
\begin{eqnarray} 
\Sigma^{(\pm)} =\frac{1}{2}(\Sigma_1\pm i\Sigma_2). \nonumber \\  
\nonumber
\end{eqnarray}
For $B_i=B\delta_{i3}$ the following identities are also useful for
implementing the above
\begin{eqnarray} 
\Sigma_3 O_\pm=\pm \frac{\hat n}{|\hat n|}O_\pm, \ 
\Sigma^{(+)}O_\pm=N_\pm \Sigma^{(+)}, \ \Sigma^{(-)}O_\pm=N_\mp \Sigma^{(-)},
\nonumber  
\end{eqnarray}
and 
\begin{eqnarray}
P_-\Sigma^{(+)}\Psi&=&\left(0,0,-\Psi^4,0\right)^{\rm T},
\nonumber\\
P_-\Sigma^{(-)}\Psi&=&\left(0,0,0,-\Psi^3\right)^{\rm T},
\nonumber\\
P_+\Sigma^{(+)}\Psi&=&\left(-\Psi^2,0,0,0\right)^{\rm T},
\nonumber\\
P_+\Sigma^{(-)}\Psi&=&\left(0,-\Psi^1,0,0\right)^{\rm T}.
\end{eqnarray}
We thus get for the self-dual case 
\begin{eqnarray}
\chi^{-+}
&=&
\left( 
\begin{array}{c}
0 \\
0  \\
\frac{1}{i\lambda}[\partial_r-\frac{\hat Br}{2}-\frac{2}{r}K_2^3]
N_-   
z^{k/2}e^{-z/2}
M\left(\frac{k}{2}+m_{2}-\Lambda^2+2,k+2,z\right)
 {\cal C}_{km_1m_2}\\
\frac{1}{i\lambda'}[\partial_r-\frac{\hat Br}{2}+\frac{2}{r}K_2^3]
N_+
z^{k'/2}e^{-z/2}
M\left(\frac{k'}{2}-m'_{2}-\Lambda^{\prime 2}+2,k'+2,z\right)
{\cal C}_{k'm'_1m'_2}
\end{array} \right)
\nonumber\\
&-&\frac{2}{r}
\left( 
\begin{array}{c}
0 \\
0  \\
\frac{1}{i\lambda'}K_2^-
N_+
z^{k'/2}e^{-z/2}
M\left(\frac{k'}{2}-m'_{2}-\Lambda^{\prime 2}+2,k'+2,z\right)
{\cal C}_{k'm'_1m'_2}
\\
\frac{1}{i\lambda}K_2^+
N_-
z^{k/2}e^{-z/2}
M\left(\frac{k}{2}+m_{2}-\Lambda^2+2,k+2,z\right)
 {\cal C}_{km_1m_2}
\end{array} \right)
\end{eqnarray}
and
\begin{eqnarray}
\chi^{--}
&=&
\left( 
\begin{array}{c}
0 \\
0  \\
\frac{1}{i\lambda}[\partial_r+\frac{\hat Br}{2}-\frac{2}{r}K_2^3]
N_+
z^{k/2}e^{-z/2}
M\left(\frac{k}{2}-m_{2}-\Lambda^2,k+2,z\right)
{\cal C}_{km_1m_2}\\
\frac{1}{i\lambda'}[\partial_r+\frac{\hat Br}{2}+\frac{2}{r}K_2^3]
N_-z^{k'/2}e^{-z/2}
M\left(\frac{k'}{2}+m'_{2}-\Lambda^{\prime 2},k'+2,z\right)
{\cal C}_{k'm'_1m'_2}
\end{array} \right)
\nonumber\\
&-&\frac{2}{r}
\left( 
\begin{array}{c}
0 \\
0  \\
\frac{1}{i\lambda'}K_2^-
N_-z^{k'/2}e^{-z/2}
M\left(\frac{k'}{2}+m'_{2}-\Lambda^{\prime 2},k'+2,z\right)
{\cal C}_{k'm'_1m'_2}\\
\frac{1}{i\lambda}K_2^+
N_+
z^{k/2}e^{-z/2}
M\left(\frac{k}{2}-m_{2}-\Lambda^2,k+2,z\right)
{\cal C}_{km_1m_2}
\end{array} \right).
\end{eqnarray}
By inspection, the boundary condition $\chi=-e^{-i \alpha} \varphi$
can only be fulfilled if terms with raising/lowering operators
of the azimuthal
quantum numbers vanish since these terms contain the
projectors $N_{\pm}$  while the rest of terms entering the boundary condition 
contain $N_{\mp}$ (see Eqs.~(\ref{b4}) and (\ref{b5})).
In particular, $m'_2=-k/2, \ m_2 = k/2$.
Finally, evaluating the derivatives of the confluent hypergeometric
functions with the help of relation
\begin{eqnarray*}
M^\prime_z(a,b,z)=\frac{a}{b} M(a+1,b+1,z) 
\end{eqnarray*}
leads to the solutions given in the main body of the paper.

\end{document}